# SOLAR SAIL PROPULSION: ENABLING NEW CAPABILITIES FOR HELIOPHYSICS

L. Johnson[1], R. Young[1], D. Alhorn[1], A. Heaton[1], T. Vansant[2], B. Campbell[2], R. Pappa[3], W. Keats[3], P. C. Liewer[4], D. Alexander[5], J. Ayon[4], G. Wawrzyniak[6], R. Burton[7], D. Carroll[7], G. Matloff[8], and R. Ya. Kezerashvili[8]

[1] NASA Marshall Space Flight Center, Huntsville, AL
[2] NASA Goddard Space Flight Center, Greenbelt, MD
[3] NASA Langley Research Center, Hampton, VA
[4] Jet Propulsion Laboratory, California Institute of Technology, Pasadena, CA
[5] Rice University, Houston, TX
[6] Purdue University, West Lafayette, IN
[7] CU Aerospace, Champaign, IL
8 New York City College of Technology, CUNY, New York, NY

**ABSTRACT**

Solar sails can play a critical role in enabling solar and heliophysics missions. Solar sail technology within NASA is currently at 80% of TRL-6, suitable for an in-flight technology demonstration. It is conceivable that an initial demonstration could carry scientific payloads that, depending on the type of mission, are commensurate with the goals of the three study panels of the 2010 Heliophysics Survey. Follow-on solar sail missions, leveraging advances in solar sail technology to support Heliophysics Survey goals, would then be feasible. This white paper reports on a sampling of missions enabled by solar sails, the current state of the technology, and what funding is required to advance the current state of technology such that solar sails can enable these missions.

**INTRODUCTION**

Solar sail propulsion uses sunlight to propel vehicles through space by reflecting solar photons from a large (> 100 meters per side), mirror-like sail made of a lightweight, highly reflective material. The continuous photonic pressure provides propellantless thrust to perform a wide range of advanced maneuvers, such as to hover indefinitely at points in space, or conduct orbital plane changes more efficiently than conventional chemical propulsion. Eventually, a solar sail propulsion system could propel a space vehicle to tremendous speeds—theoretically much faster than any present-day propulsion system. Since the Sun supplies the necessary propulsive energy solar sails require no onboard propellant, thereby significantly increasing useful payload mass.

Practical concepts for solar sailing have existed for approximately 100 years, beginning with Tsiolkovsky and Tsander in the 1920s. A team at JPL completed the first serious mission study in the late 1970s for a rendezvous with Halley's Comet (Friedman, 1978). An effort by McInnes in the 1990s and the publication of his PhD dissertation as a textbook on solar sailing helped re-invigorate interest in solar sailing as a research topic (McInnes, 1999). In the early-to-mid 2000's, NASA's In-Space Propulsion Technology Project made substantial progress in the development of solar sail propulsion systems. Two different 20-m solar sail systems were produced and successfully completed functional vacuum testing in the Glenn Research Center's (GRC's) Space Power Facility at Plum Brook Station, Ohio (Johnson, 2007). Solar sail propulsion was then selected as a candidate for flight validation by NASA's New Millennium Program on their proposed Space Technology 9 (ST-9) flight. Unfortunately, funding for both the In-Space Propulsion Technology Project and the New Millennium Program were terminated shortly thereafter due to NASA's changing priorities toward Project Constellation.

Outside of NASA, solar sailing has been tested in space. In the summer of 2010, the Japanese Aerospace Exploration Agency, JAXA, launched a solar sail spacecraft named IKAROS in tandem with another mission to Venus. The sailcraft IKAROS (14 m by 14 m) is the first in-flight demonstration of solar sailing (Tsuda, 2010). While the effects of solar radiation pressure (SRP) are smaller on this sailcraft as compared to other concepts for solar sails, numerous program objectives have been achieved, including verifying solar radiation pressure (SRP) effects on the sail and performing in-flight guidance and navigation techniques using the solar sail.

Two programs based in the United States aim to deploy their own solar sails in space before the end 2010.

NASA's *NanoSail-D*, slated to launch November 2010, has two mission requirements: (1) to successfully stow and deploy the sail and (2) to demonstrate deorbit functionality. The Planetary Society's *Lightsail-1* (5.65 m by 5.65 m), is of similar configuration to *NanoSail-D*. *LightSail-1* is a combination of three 10 cm cubesats intended to demonstrate deployment in low Earth orbit. Funded by NASA's Small Business Innovative Research, CU Aerospace and The University of Illinois are developing the *CubeSail* (Burton et al, 2010), which consists of two nearly-identical cubesat satellites to deploy a 250-m long, 20m$^2$ sail. The design can be scaled to build a heliogyro solar sail that can potentially achieve square kilometer sail areas.. The *CubeSail* is not yet scheduled for flight.

Because of the continuous force provided by solar radiation pressure on a solar sail, solar sail spacecraft can fly in non-Keplerian orbits and can continually maneuver throughout flight without the use of finite propellant. The first section of this white paper identifies key solar and heliospheric missions proposed by in recent years that are enabled by solar sail propulsion. The next section provides more details on the current state of solar sailing technology and what improvements are required to advance solar sail technology to levels necessary to accomplish sail-enabled solar and heliospheric missions. The paper concludes with funding recommendations to achieve those necessary advances.

**SCIENTIFIC MISSIONS ENABLED BY SOLAR SAILS**

A number of science mission concepts have been identified that make optimum use of solar sail technology as the next phase in the development of solar sail propulsion as the go-to technology for high C3 missions. We briefly highlight four concept missions leaving the main details to the cited references. These are Solar Polar Imager (see also white paper by Liewer et al.), GeoSail (McInnes et al. 2001), Heliostorm, and Interstellar Probe (Mewaldt et al 2001). Each of these mission concepts have been envisaged assuming the successful development of the relevant technological capability and have significantly detailed science goals of relevance to heliophysics and that are not generally achievable without access to reliable solar sail propulsion and control. Table 1 summarizes the key parameters of these missions.

**Table 1. Solar sail mission concept designs.**

| Mission | Sail dimension | Sail areal density | Orbital/Trajectory parameters | Primary Science Objective |
|---|---|---|---|---|
| **Solar Polar Imager** | 150 – 180 m on a side | 8.5-14 gm$^{-2}$ | Heliocentric with semi-major axis of 0.48 AU at inclination of 75 degrees | Convective flows in polar regions and the deep solar interior |
| **Heliostorm** | 100m x 100m | ~15 gm$^{-2}$ | Sun-centered orbit at artificial Lagrange point (0.967 AU from Sun) | Advanced Early Warning of Geo-effective solar events |
| **GeoSail** | 41.2m x 41.2m | ~33 gm$^{-2}$ | Earth Orbit: 11$R_E$ Perigee, 23 $R_E$ Apogee; Period of 4.5 days | Extended presence in Earth's geotail using sail to precess inertial orbit |
| **Interstellar Probe** | ~400m diameter | ~1 gm$^{-2}$ | Solar swing by (0.25 AU) then radial trajectory at ~15AU/yr (sail jettisoned at 5 AU | Measurements of the nearby interstellar medium |

A concept that uses a hybrid low-thrust and solar sail propulsion system has been proposed to orbit above one of Earth's poles, providing a continual view of that pole. This Earth Polesitter mission could be used to continually monitor polar weather, observe the interaction between solar wind and the Earth's magnetic field in the Earth's atmosphere, and perform geological surveys at high latitudes (Ceriotti and McInnes, 2010).

Additionally, the observation of the trajectory of a solar sail by itself may provide a test of the fundamental principles of general relativity. The curvature of spacetime, in conjunction with solar radiation pressure, affects the bound orbital motion of solar sails and leads to deviations from Kepler's third law for heliocentric and non-Keplerian orbits, as well as to the new phenomenon for non-Keplerian orbits when the orbital plane precesses around the sun that is an analog of the Lense-Thirring effect (Kezerashvili and Vazquaz-Poritz, 2009; 2010).

**CURRENT TECHNOLOGY AND NEAR-FUTURE IMPROVEMENTS**

NASA's In-Space Propulsion Technology Project funded the development of two prototype solar sail systems for ground testing. These prototypes were used to validate design concepts for sail manufacturing, packaging, launch and deployment in space; attitude control subsystem function; and characterization of the structural mechanics and dynamics of the deployed sail in a simulated space environment. Two square sail architectures were developed. They consisted of 4-quadrant, reflective sail membranes, a deployable sail support structure, an attitude control subsystem, and hardware necessary to stow the sail for launch. Additionally, these systems were required to meet the characteristics given in Table 2. A sub-L1 solar monitoring mission concept was provided as a reference mission for guidance in design and scalability issues.

NASA awarded two ground demonstration contracts. ATK Space Systems' approach incorporated their rigid coilable boom, a sliding mass attitude control system (ACS) subsystem, and partner SRS' CP1™ sail membrane (Murphy, 2005). L'Garde Inc. used their inflatable and cold-temperature rigidizable boom, a control vane based ACS, and Mylar for the sail membrane (Lichodziejewski, 2006). The parallel testing and development of these two system-level demonstrations that have varied technologies in the three major components removed the risk to this technology development if one provider encountered an unrecoverable failure. The system-level ground demonstration work was divided into three phases. A 6-month concept refinement phase was completed in May 2003. During this phase, the two teams provided analysis of their system's performance when scaled to the Design Reference Mission, and a preliminary test plan for the following two 12-month phases. The first 12-month hardware development phase began in June 2003. In this phase both teams built and tested components and subsystems, with ATK concentrating on a single 10-m quadrant, and L'Garde developing a 10-m square sail. The most comprehensive of these tests occurred in mid-2004 when the respective teams deployed their integrated subsystem in the NASA Langley Research Center (LaRC) 14-m vacuum facility (ATK), and the 30-m vacuum chamber at GRC's Plum Brook Space Power Facility (L'Garde). Following this phase, the teams culminated their work in a 12-month system verification phase. In this phase both teams built and tested fully integrated 20-m sail systems that included a launch packaging container and operational ACS subsystems. In mid-2005, the respective teams tested their systems in the Plum Brook Facility under a high vacuum and appropriate thermal environment, and subjected their systems to launch vibration and ascent vent tests. Figure 1 shows the 20-m deployed systems at the Plum Brook Facility. Since the sails represent the largest systems that could be tested in a vacuum chamber on the ground, a significant effort was made to collect static and dynamic data on the sails and booms with ~400 Gb of data collected, primarily raw photogrammetric data.

**Table 2. Characteristics of system-level solar sail ground demonstrations**

| Metric | RFP | ATK | L'Garde |
|---|---|---|---|
| **Dimensions** | 20 meters × 20 m or greater | • 20-m system with flight like central structure<br>• 4 sails scaled from 80 m<br>• Truncated 80-m masts<br>• Central structure scaled from 40 m | • 19.5 m due to Plumbrook<br>• 1 subscale TVCAD vane<br>• No flight central structure scaled for 100-m system<br>• Sails and mast truncated 100-m system |
| **Sail Subsystem Areal Density** | $< 20$ g/m$^2$ (scalability to 12 g/m$^2$ for $10^4$ m$^2$) | • 112 g/m$^2$ — includes spacecraft bus structure, ACS, power, instrument boom<br>• Scaled to 11.3 g/m$^2$ for 100-m design and no payload | • 30 g/m$^2$ – includes ACS (4 vanes calculated), central structure dropped<br>• Scaled to 14.1 g/m$^2$ with 50-kg payload and 41.4-kg bus |
| **Stowed Volume** | < 0.5 m$^3$ (scalability to 1.5 m$^3$ for 104 m$^2$) | • 0.9 m$^3$ scaled to 1.5 m$^3$ for 100-m design | • 2.14 m$^3$ scaled to 1.04 m$^3$ for 100-m design |
| **Thrust Vector Turning Rate About Roll Axis** | > 1.5°/hr | • > 35° maneuver in 2 hrs | • 63°/hr (.0175°/sec) |
| **Effective Sail Reflectance** | > 0.75 | • 92% over solar spectrum | • 85.9 |
| **Anti-sunward Emissivity** | > 0.30 | • 0.30 for 3-μ film | • 0.40 |
| **Membrane Characteristics** | Space-durable, tear-resistant, designed for 1 yr in the near-GEO environment | • ~2-μm CP1™ with 1,000 Å of aluminum on front, uncoated CP1™ on back of sail. All materials have space flight heritage | • 2-μm Mylar☐ with 1,000 Å of aluminum on front and 200 A black chromium on back |
| **System Flatness** | Effective for propulsion | • 3-point quadrant support with shear compliant border to insure a flat sail, with a proper stress level to obtain local flatness | • Striped net loss ~ 2 % |
| **ACS** | 3-axis, minimize propellant usage | • Primary: Sliding trim control masses on truss for pitch and yaw control and rotating tip bars for pinwheel control of roll. Secondary: Micro PPT thrusters at boom tips. | • Primary: propellant-less using four tip vanes |

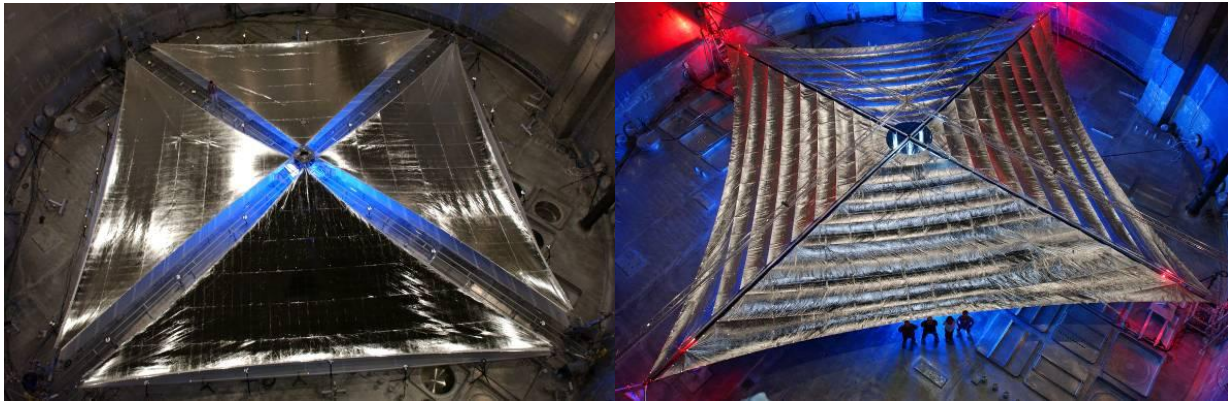

**Fig. 1. Two 20m x 20m solar sail prototypes are shown — ATK (left) and L'Garde (right).**

Integrated Software Tools: The In-Space Propulsion Technology Project also funded the development of two sets of integrated simulation tools to predict the trajectory, maneuvers, and propulsive performance of a solar sail during a representative flight profile, one by Princeton Satellite Systems (Thomas et al., 2004) and another by the Jet Propulsion Laboratory (Ellis et al., 2004). The tools were designed to be integrated into an optimal guidance and navigational control (GNC) subsystem on a future flight mission. The tools incorporated the following analytical models: 1) Solar radiation pressure acting on the sail as a function of sail orientation and distance from the Sun, 2) Disturbance forces acting on the sail such as gravitational torques and thermal deformation of the support structure, 3) Orbital mechanics, 4) Sail structural dynamics, 5) Attitude control system dynamics, and 6) Navigational sensors.

As part of the Ground System Demonstrations, LaRC, ATK, and L'Garde created Finite Element Models (FEMs) of the 10-m systems using both commercial off-the-shelf (COTS) and custom software. The purpose of this task was to create methods/algorithms/techniques that improved the conventional FEM of the membranes, booms, and other subsystems. Several methods were identified, but the option phases to complete and validate them were not funded, due to the success of the prime contractors in modeling the 20-m systems with conventional techniques, thus lowering the priority of this task below the level of the available funding.

Optical Diagnostic System: The requirements for this task included the development of methods for observation of the sail deployment, and monitoring of the health and integrity of the sail during and after deployment. During operation, the ODS would be available to provide shape and vibration measurements adequate enough to infer the stress state of the solar sail by aid of computational structural models, which could then feed real time into a closed loop spacecraft Guidance, Navigation and Control (GNC) system. The difficulty of photogrammetry was not insurmountable, but it did indicate that the challenges identified in the conceptual design phase were real. Combined with a better understanding of the lack of a need for on-orbit photogrammetry, the further development of a flight ODS was not pursued.

Structural Analysis and Synthesis Tools: This task also addressed the structural analysis issue by developing from the ground-up new and unconventional modeling techniques. Techniques considered included, Direct Transfer Function Modeling (DTFM) and Parameter Variation Processing (PVP). The overall objectives for this task were; completion of DTFM modeling/analysis methods for long booms, completion of capability to evaluate effects of imperfections, completion of PVP method for analyzing wrinkled membranes, and completion of test/analysis correlation by using existing test data. As in the above task, this effort was terminated prior to completion due to success using conventional modeling methods with the 20-m systems and a lack of funding.

Lightweight Attitude Control System: The objectives of this task were; (1) to design, integrate, and test a sail attitude control system (SACS) employing a two-axis gimbaled control boom, and (2) to develop a high-fidelity, multi-flexible body model of ATK's solar sail for the purpose of validating a thrust vector control (TVC) concept employing a two-axis gimbaled control boom. One of the major findings from this study was that the two-axis gimbaled control boom was not a mass efficient method of controlling a sail. A more efficient method was derived based on an offset mass moved along the booms by a clothesline-like apparatus to control pitch and yaw, and rotating stabilizer bars at the sail tips to pinwheel the sail quadrants for roll control. This finding led to a major redesign of the ATK 20-m hardware to accommodate the new TVC concept. An attitude determination and control block diagram were derived to present the application/integration of the inertial stellar compass with a range of ACS options from cp/cm offset to pulsed plasma micro thrusters (Thomas et al., 2005).

Characterization of Candidate Solar Sail Materials: The purpose of this task was to conduct laboratory characterization of several candidate solar sail materials. The space radiation and micrometeoroid environments for 1.0 and 0.5 astronomical units missions were defined and candidate materials were tested against these radiation and meteoroid environments. Through a series of learning tests, the sample hold-down design was optimized and a flexure test developed. Several samples of the SRS and L'Garde membrane materials were tested by subjecting them to gigarad levels of radiation in simulations of long duration solar wind mission type (Edwards et al., 2004). While some of the samples showed significant levels of degradation in mechanical strength, solar sail loading is so low, very little strength is needed.

Advanced Manufacturing Technologies: The purpose of this task was to investigate and develop an integrated approach to ultrathin film solar sail manufacturing. The focus was on; improving coating processes and technologies, developing sail seaming technologies for large monolithic sails, providing an integrated approach to membrane coating, acceptance, assembly and integration, and integrating future improvements into the process such as electrospun nanofibers for ripstop enhancement without added mass, and the addition of carbon black nanotubes to the sail backside to increase emissivity. The final results of this 2-year effort are the development of a scroll coating system, the development of coating capabilities of less than 2.5 μ, and the development of a membrane seaming system able to form monolithic sails with coatings as thin as 2.5 μ.

Smart Adaptive Structures: In order to mature the TRL of solar sail propulsion, advancements must be made in the pointing and dynamical control of these large space structures. This tasks' objective was to develop and verify structural analytical models, develop structural scaling laws, and develop adaptive control laws for solar sails to be verified on a >30-m vertically supported boom.

Sail Charging: Due to two extreme characteristics of future solar sail missions, the large surface area of the sail, and the long duration of potential missions, typical spacecraft charging issues will be exacerbated. The purpose of this task was to characterize charged particle environments for analyzing solar sail charging in the solar wind and at geostationary orbit, and to model surface and internal electric fields and potentials for solar sails using existing spacecraft charging models. Solar sail materials were tested in simulated charging environments to determine permeability and charge retention properties. A significant finding was that there will be little charging of the sail surfaces, ~10 v as a worst case in sunlight. The study found that problems arise if the sail material backing is non-conductive or electrically decoupled from the front surface. In that case, the shadowed back surface can reach potentials of –30 to –40 v relative to the space plasma in the solar wind on the order of arcing onset potentials. The solution is to make sure the sail material is conductive front to back and end-to-end if the sail is to be in geosynchronous orbit or in the auroral zone, and be careful with electrically isolated objects in the shadow of the sail.

Long-Term Space Environmental Effects: Critical to the development of solar sails is an investigation of space environmental effects on these large thin film materials and the edge support technologies. This task was related to the "Characterization of Solar Sail Materials" task above that used accelerated dose levels over a shorter period of time to simulate the total dose of radiation received by a material for many years. The purpose of this task is to provide critical thermal, optical, mechanical, and surface data on large sails taking into account edge stresses and edge support technologies that can only be characterized using large size sails but not at accelerated levels. These resulting test data could be used to validate the accelerated dose test methodology regarding the durability of candidate sail material (embrittlement, optical, mechanical, surface, and thermal properties).

**FUNDING RECOMMENDATIONS**

Taking solar sail propulsion technology from ground testing to flight validation and mission implementation is one of the top ten priorities described in the NASA Office of Chief Technologist's draft In-Space Propulsion Systems Technology Roadmap. The next step in making this technology available for use in future missions is flight validation of a full-scale (> 1000m$^2$) solar sail propulsion system. Based on the NASA ST-9 mission concept study, the development cost for such a system would be less than $200M (not including the cost of launch or mission operations). In addition to funding for solar sail technology programs, we also advocate funding on the order of $250K to $500K per year (5 to 10 grants per year) to be distributed to NASA centers and universities for solar sail mission design. Solar sail trajectory analysis leads to new mission applications, uncovering new regimes for heliophysics scientific exploration. This grant program would stimulate new research into solar sail missions and encourage growth and development of new methods and mission analysis tools for solar sail trajectory, navigation, and attitude control analysis. Assuming 3-year grants, the total program cost for this additional research would be on the order of $0.75 to 1.5M.

**CONCLUSIONS**

The TRL of solar sailing technology continues to advance. It has been confirmed by flight testing in deep space that Solar Sail technology is viable for space flight operations. Many important applications of solar sails have been identified as useful to the international science community, especially missions significant to the goals of the 2010 Solar and Heliophysics Decadal Survey. Funding for solar sail technology advancement, including continued examination of solar sail mission applications, is highly recommended as an outcome of the 2010 Decadal Survey.